\documentclass[aps,prl,graphicx,twocolumn,showpacs,preprintnumbers,amsmath,amssymb,superscriptaddress]{revtex4-1}

\usepackage{graphicx}

\begin{document}

\title{Magnetoresistance and Kondo effect in the nodal-line semimetal VAs$_2$}

\author{Shuijin Chen}
\author{Zhefeng Lou}
\author{Qin Chen}
\author{Binjie Xu}
\author{Chunxiang Wu}
\affiliation{Department of Physics, Zhejiang University, Hangzhou $310027$, China}
\author{Jianhua Du}
\affiliation{Department of Applied Physics, China Jiliang University, Hangzhou $310018$, China}
\author{Jinhu Yang}
\affiliation{Department of Physics, Hangzhou Normal University, Hangzhou $310036$, China}
\author{Hangdong Wang}
\affiliation{Department of Physics, Hangzhou Normal University, Hangzhou $310036$, China}
\author{Minghu Fang}\email{Corresponding author: mhfang@zju.edu.cn}
\affiliation{Department of Physics, Zhejiang University, Hangzhou $310027$, China}
\affiliation{Collaborative Innovation Center of Advanced Microstructures, Nanjing University, Nanjing $210093$, China}

\date{\today}

\begin{abstract}
We performed calculations of the electronic band structure and the Fermi surface as well as measured the longitudinal resistivity $\rho_{xx}(T,H)$, Hall resistivity $\rho_{xy}(T,H)$, and magnetic susceptibility as a function of temperature and various magnetic fields for VAs$_2$ with a monoclinic crystal structure. The band structure calculations show that VAs$_2$ is a nodal-line semimetal when spin-orbit coupling is ignored. The emergence of a minimum at around 11 K in $\rho_{xx}(T)$ measured at $H$ = 0 demonstrates that an additional magnetic impurity (V$^{4+}$, $S$ = 1/2) occurs in VAs$_2$ single crystals, evidenced by both the fitting of $\rho_{xx}(T)$ data and the susceptibility measurements. It was found that a large positive magnetoresistance (MR) reaching 649\% at 10 K and 9 T, its nearly quadratic field dependence, and a field-induced up-turn behavior of $\rho_{xx}(T)$ emerge also in VAs$_2$, although MR is not so large due to the existence of additional scattering compared with other topological nontrival/trival semimetals. The observed properties are attributed to a perfect charge-carrier compensation, which is evidenced by both calculations relying on the Fermi surface and the Hall resistivity measurements. These results indicate that the compounds containing V ($3d^3 4s^2$) element as a platform for studying the influence of magnetic impurities to the topological properties.
\end{abstract}

\maketitle

Since the discovery of quantum Hall effect in the two-dimensional electron gas in the 1980s \cite{PhysRevLett.50.1395}, it has been recognized that the topology of band structure in the solid materials plays an important role in the classification of maters and the understanding of physical properties. Then, many topological materials, such as, topological insulators \cite{PhysRevLett.95.226801,PhysRevLett.96.106802,PhysRevLett.103.266801}, Dirac semimetals \cite{Liu864,PhysRevB.88.125427}, Weyl semimetals \cite{shekhar2015extremely,PhysRevX.5.031013,du2016large} and nodal-line semimetals \cite{PhysRevB.99.045143,PhysRevLett.117.016602,PhysRevB.95.245113,PhysRevB.102.165133} have been proposed theoretically and confirmed experimentally. Most of them are free of strong correlation effects. In the presence of strong electron interactions, very fruitful topological phases can be expected, such as the topological Mott \cite{pesin2010mott} or Kondo insulators \cite{PhysRevLett.104.106408,PhysRevB.85.045130,PhysRevLett.110.096401}, topological superconductors \cite{RevModPhys.83.1057}, and fractional topological insulators \cite{2011Fractional,PhysRevX.1.021014,PhysRevLett.106.236804}. In order to pursue these exotic phases, searching for suitable compounds, which are strongly correlated (usually in $d$ and $f$ orbital systems) and topologically nontrivial, attracts much attention. For instance, SmB$_6$, as a typical rare-earth mixed valence compound, in which, has been proposed theoretically as a topological Kondo insulator \cite{PhysRevLett.104.106408,PhysRevB.85.045130,PhysRevLett.110.096401} and recently has been confirmed by transport \cite{PhysRevB.88.180405,kim2013surface,2014Two,PhysRevB.94.205114}, photoemission \cite{PhysRevB.88.121102,2013Observation,denlinger2014smb6} and scanning tunneling microscope (STM) \cite{yee2013imaging} experiments. At sufficiently low temperature, the hybridization between $4f$ orbitals and highly dispersive $5d$ bands results in the formation of ``heavy fermion'' bands, thus, SmB$_6$ is a correlated $\mathcal{Z}_2$ topological insulator. Recently, a number of groups \cite{autes2019topomat,vergniory2019complete,zhang2019catalogue,tang2019comprehensive} predicted thousands of candidate topological materials by performing systematic high-throughput computational screening across the databases of known materials. Following these predictions, it is possible to search for the topological materials in the known compounds containing $3d$ elements based on the calculations of band structure, and the measurements of physical properties.

VAs$_2$ crystalizes in a monoclinic structure with space group C$_{12/m1}$ (No. 12), as shown in Fig. 1(a), has an isostructure to the transition metal dipnictides XPn$_2$ (X = Nb, Ta; Pn = P, As, Sb) \cite{PhysRevB.93.195106,PhysRevB.93.195119,doi:10.1063/1.4940924,PhysRevB.94.041103,PhysRevB.94.121115,PhysRevB.93.184405,TaAs2}, which have been widely studied theoretically and experimentally as a class of topological materials. For example, Baokai Wang $et$ $al.$ \cite{PhysRevB.100.205118} identified the presence of a rotational-symmetry-protected topological crystalline insulator (TCI) states in these compounds based on first-principles calculations combined with a symmetry analysis. It was found that all these compounds exhibit high mobilities and extremely large positive MR \cite{PhysRevB.93.195119,doi:10.1063/1.4940924,PhysRevB.94.041103,PhysRevB.94.121115,PhysRevB.93.184405,TaAs2}. Interestingly, a negative longitudinal MR when the applied field is parallel to the current direction was observed in both TaSb$_2$ \cite{PhysRevB.94.121115} and TaAs$_2$ \cite{TaAs2}, similar to that observed in the known weyl semimetals TaAs family \cite{PhysRevX.5.031013,PhysRevB.93.121112,shekhar2015extremely,PhysRevB.92.115428,PhysRevB.92.235104,ghimire2015magnetotransport,du2016large,hu2016pi,Xu613}. Compared to the $4d$/$5d$ electrons in NbAs$_2$/TaAs$_2$, the more localization of the $3d$ electrons in VAs$_2$ might lead to a stronger electronic correlation, or introduce an additional magnetic scattering to the carriers (Kondo effect).

In this paper, we grew successfully VAs$_2$ crystals with a monoclinic structure and measured its longitudinal resistivity $\rho_{xx}(T,H)$, Hall resistivity $\rho_{xy}(T,H)$, and magnetic susceptibility as a function of temperature at various magnetic fields, as well as calculated its electronic band structure and Fermi surface (FS). The band structure calculations show that VAs$_2$ is also a nodal-line semimetal when spin-orbit coupling (SOC) is ignored. It was found that the $\rho_{xx}(T)$ measured at $H$ = 0 exhibits a minimum at around 11 K, which is considered of originate from the magnetic impurities V$^{4+}$ ($S$ = 1/2) scattering ($i.e.$ Kondo effect) , evidenced by both the susceptibility measurements and the fitting of $\rho_{xx}(T)$ data at low temperatures. We further reveal a nearly quadratic field dependence of MR, reaching 649\% at 10 K and 9 T, a field-induced up-turn behavior of $\rho_{xx}(T)$ in this material, which are attributed to a perfect charge-carrier compensation, evidenced by both the calculations relying on the FS topology and the Hall resistivity measurements.

\begin{figure*}
 \centering
 \includegraphics[width=16cm]{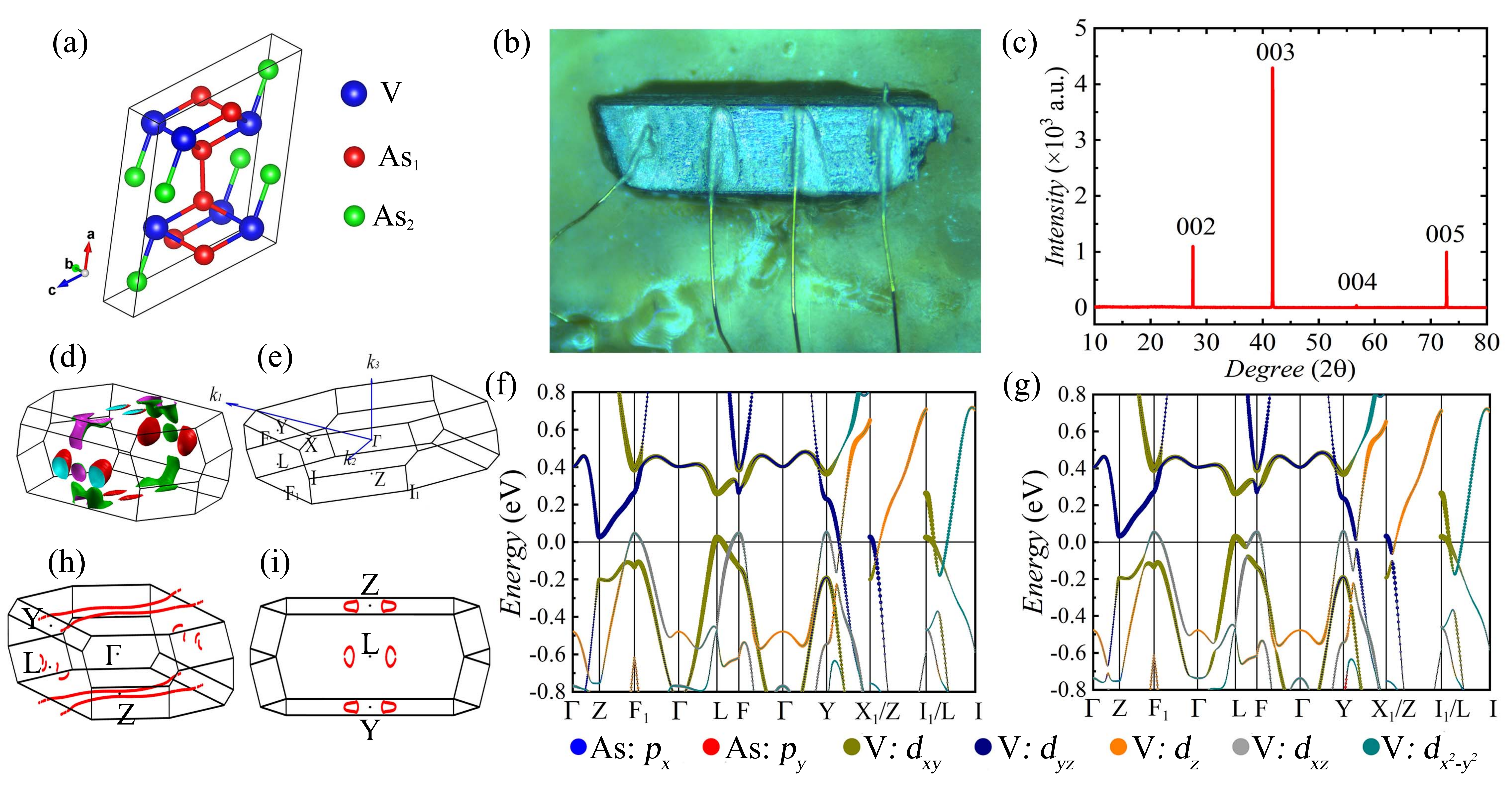}
 \caption{(a) Crystal structure of  VAs$_2$. (b) Photograph of a VAs$_2$ crystal with 4 wires for resistance measurements. (c) Single crystal XRD pattern of VAs$_2$. (d) The calculated Fermi surface of VAs$_2$. (e) The Brillouin zone. (f) and (g)  Band structures calculated without and with considering SOC. (h) and (i) Nodal lines in the first Brillouin zone from different perspective. }
\end{figure*}

\begin{figure}
\centering
\includegraphics[width=8cm]{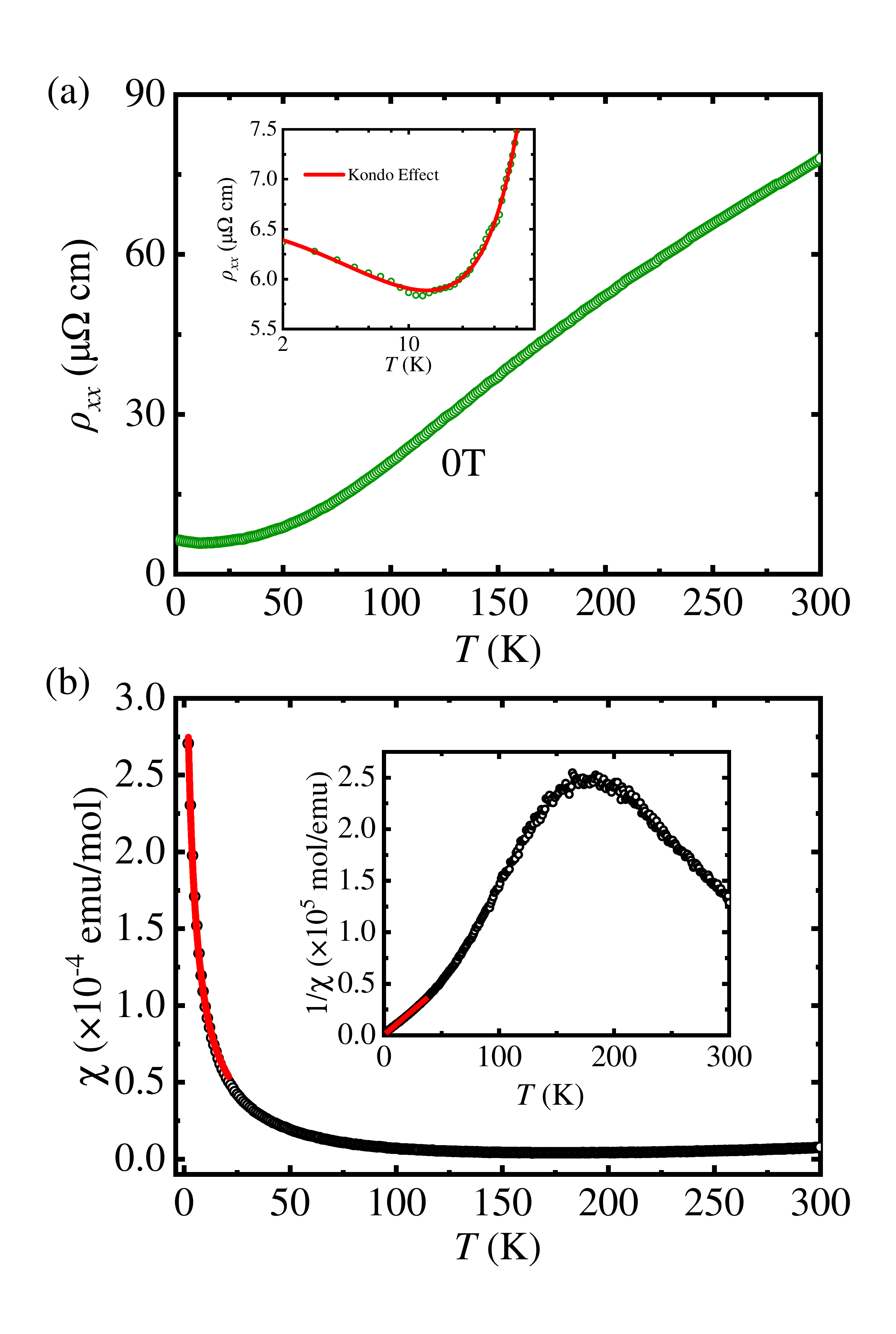}
\caption{ (a) Temperature dependence of resistivity $\rho (T)$, inset: the $\rho (T)$ data below 40 K and the fitting by Eq. (2). (b) magnetic susceptibility, measured at 1 T, inset: 1/$\chi$(\emph{T}). }
\end{figure}

\begin{figure}
\centering
\includegraphics[width=8cm]{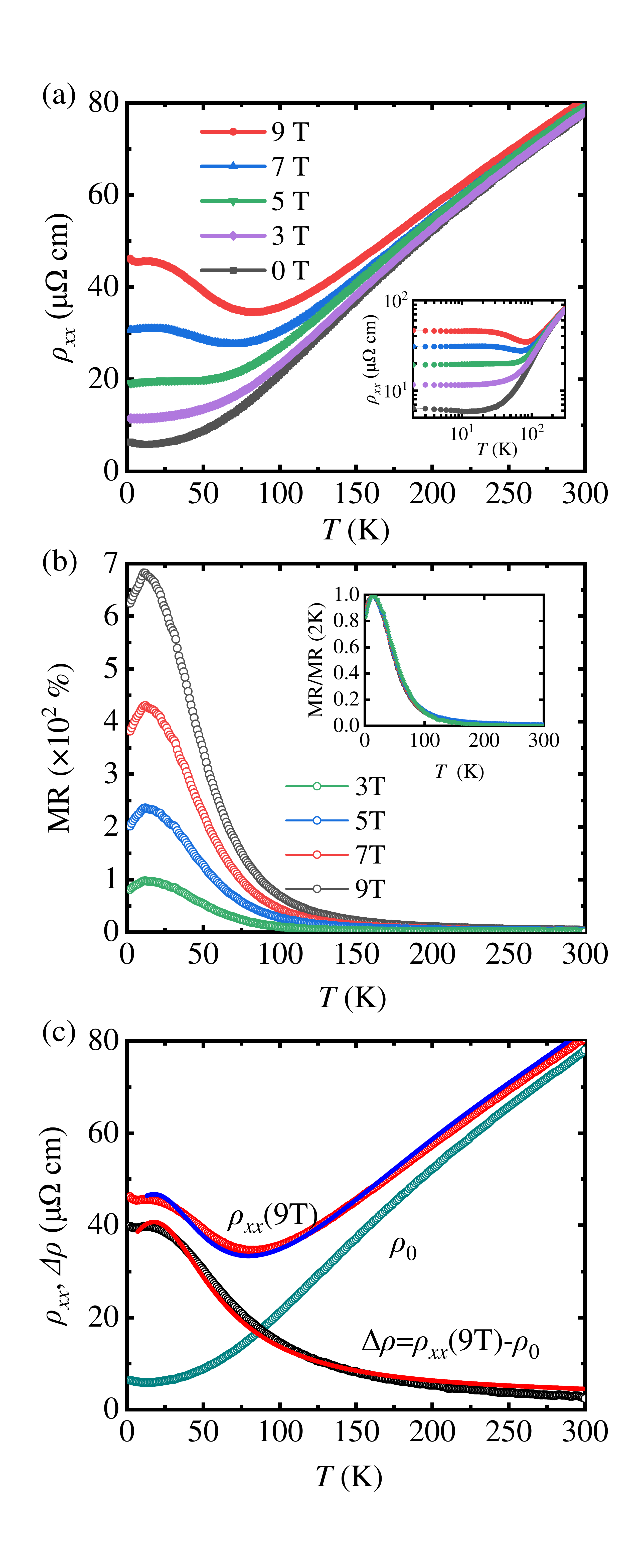}
\caption{(Color online) (a) Temperature dependence of resistivity measured at various magnetic fields. (b) The  MR vs. temperature under various magnetic fields. The inset is normalized MR. (c) Temperature dependence of resistivity at 0 and 9 T and their
difference. The red and blue lines are the fitting lines using the Kohler scaling law. }
 \end{figure}

 \begin{figure}
\centering
\includegraphics[width=8cm]{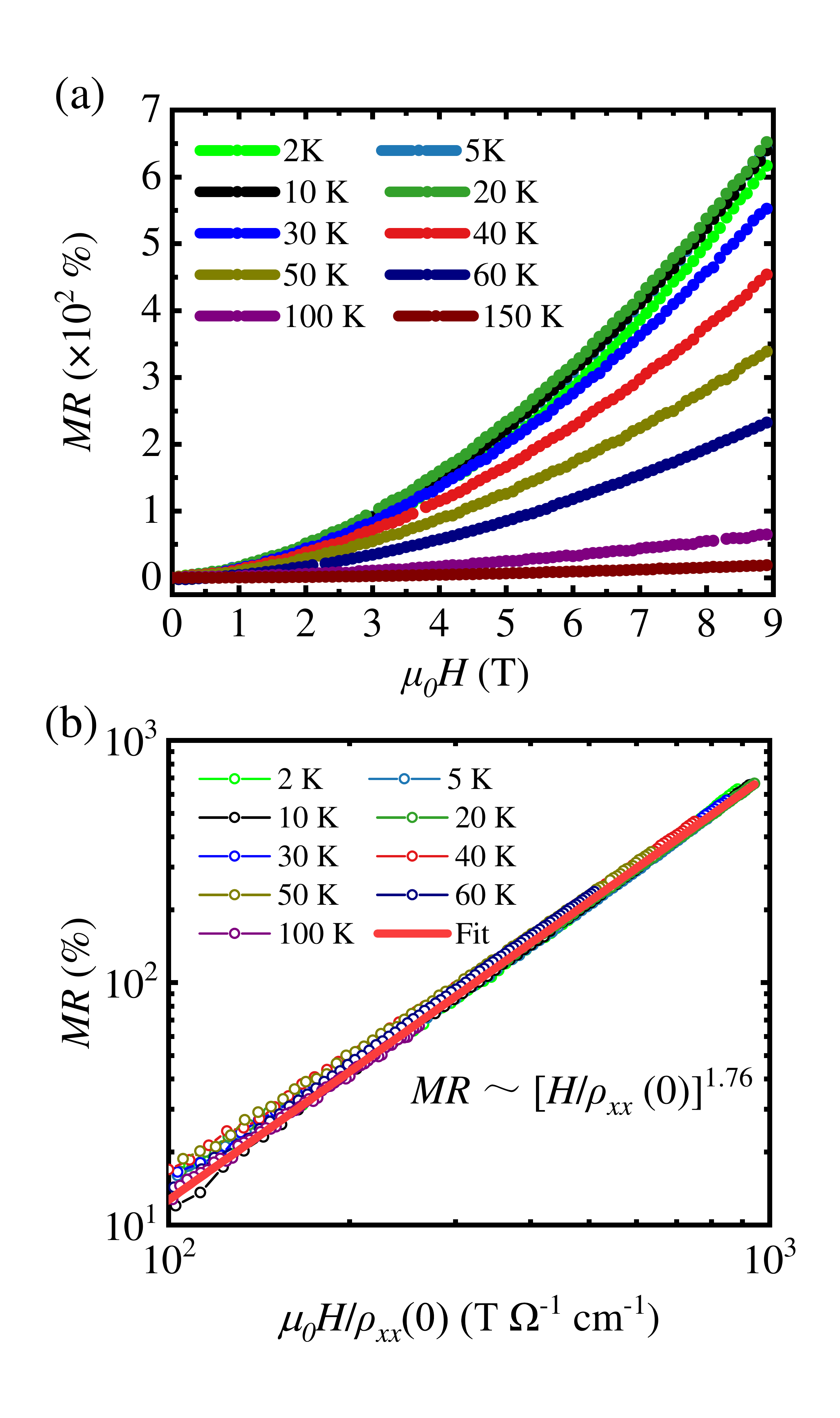}
\caption{ (a) Field dependence of MR at various temperature. (b) MR plotted as a log scale as a function of \emph{H}/$\rho$$_{xx}$(0). }
 \end{figure}

\begin{table*}
  \renewcommand\arraystretch{1.6}
  \centering
  \caption{The obtained parameters by the fitting to $\rho(T)$ data using Eq. (2).}
  \setlength{\tabcolsep}{5mm}
  {
  \begin{tabular}{ccccccc}
    \toprule
    $\rho_0$ ($\mu\Omega$ cm) & $\rho_b$ ($\mu\Omega$ cm) & $b$ ($\mu\Omega$ cm/K$^{2}$) & $c$ ($\mu\Omega$ cm/K$^{5}$) & $T_{K}$ (K)& $T_{W}$ (K) & $S$ \\
    \hline
    2.33 & 4.7 & 1.3$\times$10$^{-3}$ & 1.0$\times$10$^{-9}$ & 7.23 & 0.78 & 0.5 \\
    \botrule
  \end{tabular}
  }
\end{table*}

 \begin{figure*}
\centering
\includegraphics[width=16cm]{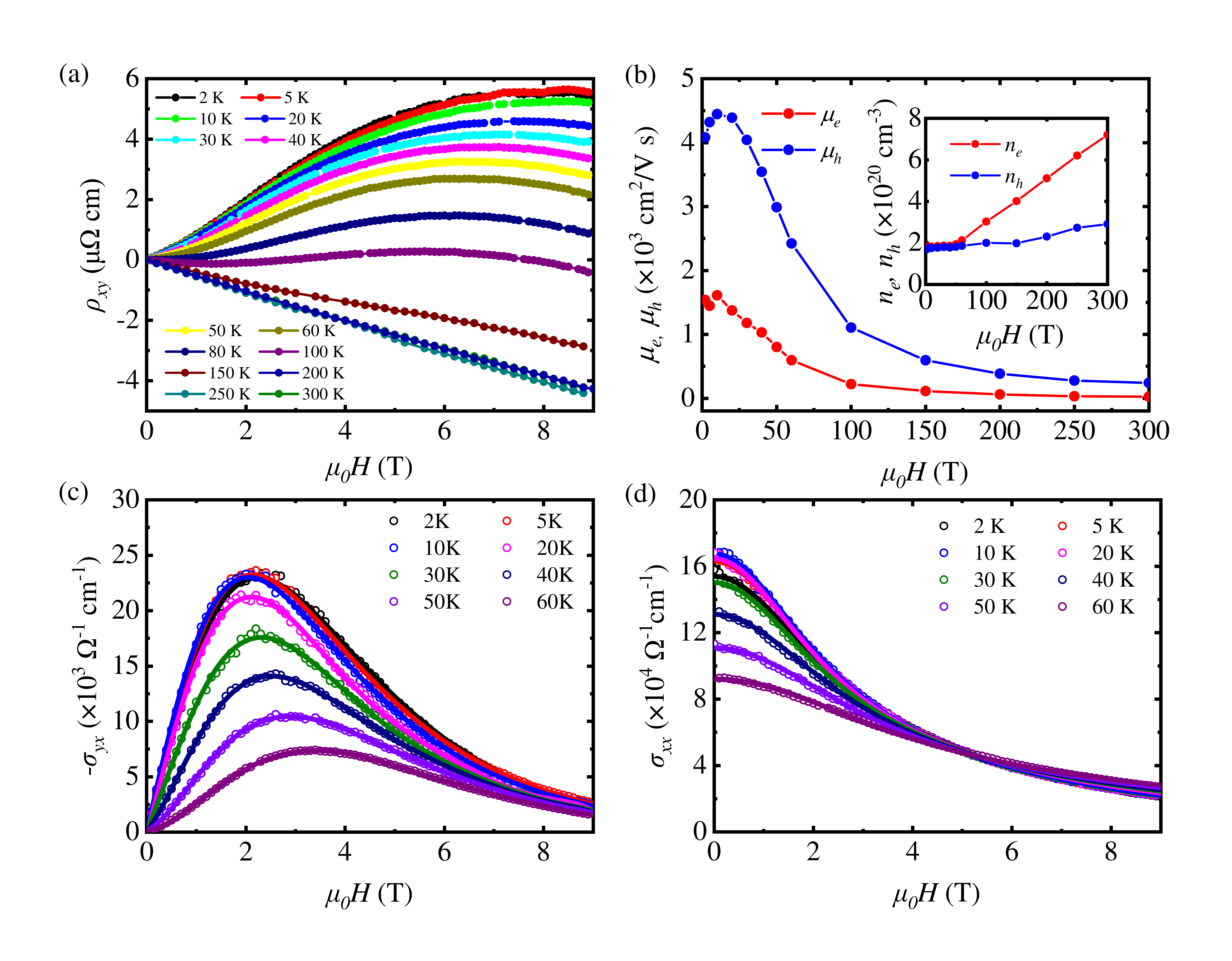}
\caption{ (a) Field dependence of Hall resistivity $\rho_{xy}$ measured at various temperatures for VAs$_2$ crystal. (b) Charge-carrier mobilities $\mu_e$ and $\mu_h$, and (inset) the carrier concentrations, $n_e$ and $n_h$, as a function of temperature extracted from the two-carrier model analysis of both $\sigma_{xy}$ and $\sigma_{xx}$ data. Components of the conductivity tenser, $i.e.$ $\sigma_{xy}$ and $\sigma_{xx}$ in the panels (c) and (d), respectively, as a function of magnetic field for different temperatures ($<$ 60 K). Hollow dots represent experimental data and solid lines are the fitting curves by using the two-carrier model. }
\end{figure*}

VAs$_2$ single crystals were grown by a chemical vapor transport method. High purity V and As powder were mixed in a mole ratio 1 : 2, then sealed in an evacuated silica tube containing I$_2$ as a transport agent with 10 mg/cm$^ 3$. The quartz tube was placed in a tube furnace with a temperature gradient of 950$^{\circ}$C - 750$^{\circ}$C and heated for two weeks. Polyhedral crystals were obtained at the cold end of the tube. A single crystal with a dimension of $2.0\times0.73\times0.29$ mm$^3$ [see Fig. 1(b)] and a cleavage surface (001) was selected for the transport and magnetic property measurements. The composition was detected to be V : As = 32.7 : 67, using the Energy Dispersive X-ray Spectrometer (EDXS), the crystal structure was determined by the single-crystal X-Ray diffraction, as shown in Fig. 1(c), from which, the lattice parameter, $c$ = 7.481(7) \AA, consistent with the result in Ref. \cite{hulliger1964new}. The longitudinal resistivity  and Hall resistivity measurements were carried out on a physical property measurement system (Quantum Design, PPMS-9 T) with a standard four-probe method [see Fig. 1(b)]. The magnetization measurements were carried out on a magnetic property measurement system (Quantum Design, MPMS-7 T). The band structure was calculated by using the Vienna $ab$ $initio$ simulation package (VASP) \cite{kresse1996efficient,kresse1999g} with a generalized gradient approximation (GGA) of Perdew, Burke and Ernzerhof (PBE) \cite{perdew1996phys} for the exchange correlation potential. A cutoff energy of 520 eV and a $10\times10\times6$ $k$-point mesh were used to perform the bulk calculations. The nodal-line search and FS calculations were performed by using the open-source software WannierTools \cite{wu2018wanniertools} that is based on the Wannier tight-binding model (WTBM) constructed using Wannier90 \cite{mostofi2014updated}.

First at all, we discuss the results of our electronic structure calculations that extend the initial prediction of the high symmetry line semimetal for VAs$_2$ \cite{zhang2019catalogue}. In order to address the topological character of VAs$_2$, we calculate its band structure and the FS. As shown in Fig. 1(d), there are six different FS sheets: two hole-like surfaces (green) at the L and Y points, and four electron-like (red) surfaces near L and Z points, the volume of the electron and hole pockets is roughly the same, indicating that VAs$_2$ is nearly an electron-hole compensated semimetal, also evidenced by the Hall resistivity measurements discussed as follows. The bulk band structure of VAs$_2$ without and with SOC is presented in Fig. 1(f) and 1(g), respectively. It can be seen clearly that the bands near the Fermi level mainly arise from the $d$ orbits of V atoms, and the valence and conduction bands cross along the Y-X$_1$, Z-I$_1$ and L-I high symmetry directions. Without SOC, in the Brillouin zones (BZ), the nodal lines can be found by using the open-source software WannierTool \cite{wu2018wanniertools}, see in Fig. 1(h). There are two type of nodal lines, one is two nonclosed spiral lines extending across the BZ through point Z, another is two nodal loops near the L point. When SOC is included, these nodal lines are gapped [see Fig. 1(g)] and lead to a band inversion along the Y-X$_1$, Z-I$_1$ and L-I high symmetry directions, driving into a topological crystalline insulator (TCI), as discussed by Baokai Wang $et$ $al.$ \cite{PhysRevB.100.205118} for the identical structure transition metal dipnitides RX$_2$ (R = Nb or Ta; X = P, As or Sb). Although the opening of a local band gap between the valence and conduction bands occurs when SOC is included, VAs$_2$ preserves its semimetal character with the presence of electron and hole pockets, similar to that in NbAs$_2$ reported in Ref. \cite{PhysRevB.100.205118}.

Next, we focus on the Kondo effect emerging in the longitudinal resistivity for VAs$_2$. Figure 2(a) shows the temperature dependence of resistivity, $\rho_{xx}(T)$. With decreasing temperature, the resistivity, $\rho_{xx}$, decreases monotonously from $\rho_{xx}$(300 K) = 78 $\mu\Omega$ cm, reaches a minimum at around 11 K, then increases a little to 6.4 $\mu\Omega$ cm at 2 K, thus the residual resistivity ratio (RRR) $\rho_{xx}$(300 K)/$\rho_{xx}$ (2 K) $\sim 12$, much smaller than that observed in the other transition metal dipnictide crystals, such as NbAs$_2$ ($\sim 75$) \cite{PhysRevB.94.041103}, TaAs$_2$ ($\sim 100$) \cite{TaAs2}. As we know, the system, containing magnetic impurities which usually scatter the conduction electrons through the $s-d$ exchange interactions, exhibits a minimum in resistivity at lower temperatures, $i.e.$ termed as the Kondo effect \cite{kastner1977kondo,PhysRevB.7.3215}. As discussed by S. Barua $et$ $al.$ for VSe$_2$ \cite{barua2017signatures}, considering the correction to the resistivity due to the Ruderman-Kittel-Kasuya-Yosida (RKKY) interactions between the paramagnetic V$^{4+}$ ions, the Kondo resistivity described by the Hamann expression is modified to:
\begin{eqnarray}
\rho_{sd} = \frac{\rho_0}{2}\left[1-\frac{\ln(T_{eff}/T_K)}{\left[\ln^2(T_{eff}/T_K)+S(S+1)\pi^2\right]^{1/2}}\right]
\end{eqnarray}
where $\rho_0$ is the unitarity limit, $T_K$ is the Kondo temperature and $S$ is the spin of magnetic impurity, the effective temperature $T_{eff}=(T^{2}+T^{2}_{W})^{1/2}$, in which $k_{B}T_{W}$ is the effective RKKY interaction strength \cite{kastner1977kondo}, replaces to $T$ in the original expression \cite{barua2017signatures}. The temperature dependence of resistivity, $\rho_{xx}(T)$, is expressed as :
\begin{eqnarray}
\rho(T) = \rho_{sd} + bT^2 + cT^5 + \rho_b
\end{eqnarray}
where \emph{bT}$^n$ term is the electron-phonon scattering contribution, $\rho_b$ is an independent resistivity. We used Eq. (2) to fit the resistivity data measured at low temperatures (2 - 40 K). The results are shown in the inset in Fig. 2(a), it is clear that Eq. (2) can well describe the $\rho(T)$ data below 40 K, and the obtained parameters in Eq. (1) and Eq. (2) from the fitting are listed in TABLE I. The existence of V$^{4+}$ ($S$ = 1/2) impurities in our VAs$_2$ crystal was confirmed by the magnetic susceptibility measurements. The temperature dependence of magnetic susceptibility, $\chi(T)$, measured at 1 T with a field-cooling (FC) process is presented in Fig. 2(b). With decreasing temperature, the susceptibility $\chi$ decreases a little, reaches a minimum at around 180 K, then increases strikingly below 100 K. No magnetic transition was observed in the whole temperature range (2 - 300 K), and the significant increase of $\chi$ in the lower temperatures was considered to originate from the contribution of V$^{4+}$ ($S$ = 1/2) impurities existing in crystal as the interstitial ions, as observed in VSe$_2$ crystals \cite{barua2017signatures}. We used the Curie-Weiss law $\chi=\frac{C}{T-\theta}$, to fit the  data below 40 K, as shown in Fig. 2(a), the Curie constant $C$ = 1.27 ($\pm$0.02)$\times$10$^{-3}$ emu K/mol, and the curie temperature $\theta$ = -2.61 ($\pm$0.08) K, were obtained. The nearly linear relationship between 1/$\chi$ and $T$ below 40 K [see the inset in Fig. 2(b)] demonstrates the reliability of the fitting. The V$^{4+}$ ($S$ = 1/2) impurity molar fraction was estimated to be of 0.34 ($\pm$0.02) $\%$, small amount impurities existing in the crystals.

Third, we discuss the magnetoresisitance (MR) occurring in the nodal-line semimetal VAs$_2$, with the presence of Kondo scattering of V$^{4+}$. Figure 3(a) presents the temperature dependence of longitudinal resistivity, $\rho_{xx}(T)$, measured at various magnetic fields $H$, with current $I$ applied in the (001) plane, and $H$ $\perp$ (001) plane. Similar to many other nontrivial and trivial topological semimetals \cite{PhysRevB.93.195119,doi:10.1063/1.4940924,PhysRevB.94.121115}, VAs$_2$ also exhibits a large MR. As shown in Fig. 3(a), an up-turn in $\rho_{xx}$(T) curves under applied magnetic field occurs at low temperatures: $\rho_{xx}$ increases with decreasing $T$ and then saturates. Figure 3(b) shows MR as a function of temperature measured at various magnetic fields, with the conventional definition $MR$=$\frac{\Delta\rho}{\rho(0)}$=$[\frac{\rho(H)-\rho(0)}{\rho(0)}]$ $\times$ 100$\%$. The normalized MR, shown in the inset of Fig. 3(b), has the same temperature dependence for various fields, excluding the suggestion of a field-induced metal-insulator transition \cite{zhao2015anisotropic,khveshchenko2001magnetic} at low temperatures, as discussed in our works addressing the topological trivial semimetal $\alpha$-WP$_2$ \cite{PhysRevB.97.245101}, for the nodal-line semimetal MoO$_2$ \cite{PhysRevB.102.165133}, and the work of Thoutam $et$ $al.$ on the type-\uppercase\expandafter{\romannumeral2} weyl semimetal WTe$_2$ \cite{PhysRevLett.115.046602}. Figure 3(c) displays $\rho_{xx}(T)$, measured at 0 and 9 T, as well as their difference $\Delta$$\rho$$_{xx}$ = $\rho$$_{xx}$(\emph{T}, 9 T) - $\rho$$_{xx}$(\emph{T}, 0 T). It is clear that the resistivity in a applied magnetic field consists of two components, $\rho_0(T)$ and $\Delta\rho_{xx}$, with opposite temperature dependencies. As discussed by us for $\alpha$-WP$_2$ \cite{PhysRevB.97.245101}, MoO$_2$ \cite{PhysRevB.102.165133} and by Thoutam $et$ $al$. for WTe$_2$ \cite{PhysRevLett.115.046602}, the resistivity can be written as :
\begin{eqnarray}
  \rho_{xx}(T,H) = \rho_{0}(T)[1+\alpha(H/\rho_{0})^m]
  \end{eqnarray}
The second term is the magnetic-field-induced resistivity $\Delta\rho_{xx}$, which follows the Kohler rule with two constants $\alpha$ and $m$. $\Delta\rho_{xx}$ is proportional to 1/$\rho_0$ (when $m$ = 2) and competes with the first term upon changing temperature, possibly giving rise to a minimum in $\rho(T)$ curves. Figure 4(a) presents MR as a function of field at various temperatures. The measured MR is large at low temperatures, reaching 649 $\%$ at 10 K and 9 T, and does not show any of saturation up to the highest field (9 T) in our measurements. As discussed above, MR can be described by the Kohler scaling law \cite{PhysRevB.92.180402,ziman2001electrons}:
\begin{eqnarray}
\textit{MR} = \frac{\Delta\rho_{xx}(T,H)}{\rho_{0}(T)} = \alpha(\textit{H}/\rho_{0})^{m}
\end{eqnarray}
As shown in Fig. 4(b), all MR data from 2 - 100 K collapse onto a single straight line in the plotted as $MR\sim H/\rho_0$ curve, and $\alpha$ = 0.038 $(\mu\Omega$ cm/T$)^{1.76}$ and $m$ = 1.76 were obtained by fitting. The nearly quadratic field dependence of MR observed in this nodal-line semimetal VAs$_2$ is attributed to the electron-hole compensation, evidenced by FS calculations mentioned above, as well as the Hall resistivity measurements discussed below, which is a common characteristics for the most topologically nontrivial and trivial semimetals \cite{PhysRevB.93.195119,doi:10.1063/1.4940924,PhysRevB.94.121115}. Figure 5(a) displays the Hall resistivity, $\rho_{xy}(H)$, measured at various temperatures for a VAs$_2$ crystal with $H$ $\parallel$ $c$ axis. The nonlinear field dependence of $\rho_{xy}(H)$ below 100 K demonstrates its semimetal characteristics, in which both electron and hole carriers coexist. Following the analysis of Ref. \cite{PhysRevB.94.235154} for $\gamma-$MoTe$_2$ \cite{PhysRevB.94.121101} and by us for MoO$_2$ \cite{PhysRevB.102.165133}, we analyze the longitudinal and Hall resistivity by using the two-carrier model. In this model, the conductivity tensor, in its complex representation, is giving by \cite{ali2014large}:
\begin{eqnarray}
\sigma = \frac{en_{e}\mu_{e}}{1+i\mu_{e}\mu_0H}+\frac{en_{h}\mu_{h}}{1+i\mu_{h}\mu_0H}
\end{eqnarray}
where $n_e$ ($n_h$) and $\mu_e$ ($\mu_h$) denote the carrier concentrations and mobilities of electrons (holes), respectively. To appropriately evaluate the carrier densities and mobilities, we calculated the Hall conductivity $\sigma_{xy}$ = -$\frac{\rho_{xy}}{\rho_{xx}^2+\rho_{xy}^2}$ and the longitudinal conductivity $\sigma_{xx}$ = $\frac{\rho_{xx}}{\rho_{xx}^2+\rho_{xy}^2}$ by using the original experimental $\rho_{xy}(H)$ and $\rho_{xx}(H)$ data. Then, we fit both $\sigma_{xy}(H)$ and $\sigma_{xx}(H)$ data by using the same fitting parameters and the field dependence given by \cite{PhysRevB.94.121101}:
\begin{eqnarray}
\sigma_{xy} = \frac{e(\mu_0H)n_{h}\mu_{h}^2}{1+\mu_{h}^2(\mu_0H)^2}-\frac{e(\mu_0H)n_{e}\mu_{e}^2}{1+\mu_{e}^2(\mu_0H)^2}
\end{eqnarray}

\begin{eqnarray}
\sigma_{xx} = \frac{en_{h}\mu_{h}}{1+\mu_{h}^2(\mu_0H)^2}+\frac{en_{e}\mu_{e}}{1+\mu_{e}^2(\mu_0H)^2}
\end{eqnarray}

Figures 5(c) and 5(d) display the fitting of both the $\sigma_{xy}(H)$ and $\sigma_{xx}(H)$ measured at $T$ = 2 - 60 K, respectively. The excellent agreement between our experimental data and the two-carrier model over a broad range of temperature, confirms the coexistence of electrons and holes in VAs$_2$. Figure 5(b) shows the obtained $n_e$, $n_h$, $\mu_e$ and $\mu_h$ values by fittings as a function of temperature. The almost same values of $n_e$ and $n_h$ below 60 K [see the inset Fig. 5(b)], such as $n_e$ = 1.77 $\times$10$^{20}$ cm$^{-3}$ and $n_h$ = 1.69 $\times$10$^{20}$ cm$^{-3}$ at 2 K, indicate that VAs$_2$ is indeed a electron-hole compensated semimetal, consistent with the above results from the calculation FS. Both electron and hole densities are estimated to be 10$^{20}$ cm$^{-3}$ in our VAs$_2$ crystal, significantly higher than those in Dirac semimetals, such as, Cd$_3$As$_2$ ($\sim$ 10$^{18}$ cm$^{-3}$ \cite{neupane2014observation}) and Na$_3$Bi ($\sim$ 10$^{17}$ cm$^{-3}$ \cite{Xiong413}), but comparable to these of another nodal-line semimetals ZrSiS ($\sim$ 10$^{20}$ cm$^{-3}$ \cite{PhysRevLett.117.016602}), and MoO$_2$ ($\sim$ 10$^{20}$ cm$^{-3}$ \cite{PhysRevB.102.165133}), demonstrating further the nodal-line characteristics of VAs$_2$. As shown in Fig. 5(b), it is clear that the hole mobility $\mu_h$ is higher than $\mu_e$ in the whole temperature range (2 - 300 K), such as, at 2 K, $\mu_h$ = 4.08 $\times$10$^3$ cm$^2$/Vs, $\mu_e$ = 1.54 $\times$10$^3$ cm$^2$/Vs, but smaller one order of magnitude than these observed by us in the nodal-line semimetal MoO$_2$ ($\sim$ 10$^4$ cm$^2$/Vs) \cite{PhysRevB.102.165133}, and both $\mu_h$ and $\mu_e$ decrease notably with increasing temperature due to the existence of phonon thermal scattering at higher temperatures. It is worth noting that both $\mu_h$ and $\mu_e$ have a little decrease below 11 K, corresponding to the Kondo scattering from V$^{4+}$ magnetic impurities mentioned above.

In summary, we calculated the electronic structure, and measured the longitudinal resistivity, Hall resistivity and magnetic
susceptibility for VAs$_2$. It was found that VAs$_2$ exhibits many common characteristics of the topological nontrivial/trivial semimetals, such as a large MR reaching 649$\%$ at 10 K and 9 T, a nearly quadratic field dependence of MR, and a field-induced up-turn behaviour in $\rho$$_{xx}$(\emph{T}). Both the FS calculations and the Hall resistivity measurements verify these properties being attributed to a perfect carrier compensation. Interestingly, the Kondo scattering due to the existence of V$^{4+}$ ($S$ = 1/2) magnetic impurities in our VAs$_2$ crystals occurs, indicating that VAs$_2$ crystal can be used to study the Kondo effect in the nodal-line semimetal.

ACKNOWLEDGEMENTS: This research is supported by the Ministry of Science and Technology of China under Grant No. 2016YFA0300402 and the National Natural Science Foundation of China (NSFC) (NSFC-12074335 and No. 11974095), the Zhejiang Natural Science Foundation (No. LY16A040012) and the Fundamental Research Funds for the Central Universities.

\bibliography{citation}
\end{document}